\documentclass[a4,usenatbib,fleqn]{mn2e}
\usepackage{mathptmx}
\usepackage{amsmath,amssymb,amsbsy}
\usepackage[pdftex]{graphicx}
\usepackage[backref=none]{hyperref}
\usepackage[usenames,dvipsnames]{color}
\usepackage{bm}
\usepackage{txfonts}
\DeclareSymbolFont{cmletters}{OML}{cmm}{m}{it}
\DeclareMathSymbol{v}{\mathalpha}{cmletters}{"76}

\definecolor{MyDarkBlue}{rgb}{0,0.1,0.7}
\hypersetup{pdfborder={0 0 0},colorlinks,breaklinks=true,
  urlcolor={blue},citecolor={MyDarkBlue},linkcolor={magenta}}

\title[QPO frequency derivative and boundary layer]
{QPO frequency derivative - frequency correlation indicates non-Keplerian boundary layer with a maximum in rotation rate}

\author[M. Ali Alpar]{M. Ali Alpar\thanks{E-mail:
alpar@sabanciuniv.edu}\\Sabanc{\i} University, Orhanl{\i}, Tuzla
34956, \.{I}stanbul, Turkey}

\date{Accepted XXX. Received YYY; in original form ZZZ}

\pubyear{2015}

\begin{document}
\label{firstpage}
\pagerange{\pageref{firstpage}--\pageref{lastpage}}
\maketitle

\begin{abstract}
The correlation between the frequency and the absolute
value of the frequency derivative of the kilohertz QPOs observed for
the first time from 4U 1636-53 is a simple consequence and indicator of the 
existence of a non-Keplerian rotation rate in the accretion disk
boundary layer. This paper interprets the observed correlation, showing that the observations provides strong evidence in support of the fundamental assumption of disk accretion models around slow rotators, that the boundary layer matches the Keplerian disk to the neutron star magnetosphere. 
\end{abstract}

\begin{keywords}
accretion, accretion disks -- stars: neutron -- X-rays: binaries -- X-rays: individual: 4U 1636-53
\end{keywords}
\section{Introduction}
Quasi-periodic oscillations (QPO) from neutron stars in low-mass
X-ray binaries (LMXB) \citep{vdKlisetal85, revvdKlis10} are likely to
contain information on the neutron star in interaction with
surrounding material.  The beat-frequency model
\citep{AlparShaham85}, interpreted the first discovered QPOs (later
called ``horizontal branch oscillations'') from the LMXB GX 5-1
\citep{vdKlisetal85} as the beat frequency between the rotation rate
of the neutron star and the Keplerian frequency at the inner edge of
the accretion disk. Disk modes in association with horizontal and
normal branch QPOs were further explored by \citet{Alparetal92,
AY97} and many other researchers (see \citet{revvdKlis10} and references therein). 
With the discovery of QPOs at kilohertz
frequencies \citep{vdKlisetal96, Strohmayeretal96, revvdKlis00}, 
the interaction between the neutron star magnetosphere and the 
accretion disk boundary layer was posed as the likely source of 
these high frequency modulations of the X-ray luminosity 
\citep{revvdKlis10, Campana00, Cui00}.
Modes in the boundary layer of a thin gaseous accretion disk with
inner radius quite close to the neutron star could be modulating the
accretion flow onto the neutron star at such high frequencies 
\citep{Wagoner99, Kato01, EA04, EPA08, AlparPsaltis08, Kato09, Kato12}. 
The rotation rates $\Omega(r)$ in the disk
boundary layer are likely to be non-Keplerian, as the boundary layer
matches the Keplerian flow in the disk to the rotation frequency
$\Omega_*$ of the neutron star and its co-rotating magnetosphere.  
LMXBs contain ``slow rotators'', with the neutron star rotation rate
$\Omega_*$ less than $\Omega_K(r_{in})$, the Keplerian rotation rate
in the inner disk beyond the boundary layer. The rotation rate
$\Omega(r)$ within the boundary layer must clearly go through a
maximum value $\Omega_{max}$, as it deviates from the Keplerian flow
in the disk to match the stellar rotation rate $\Omega_*$ at the
magnetosphere. \citet{AlparPsaltis08} noted that this deviation from
Keplerian flow in the boundary layer can significantly modify
constraints on the neutron star mass and radius obtained by
associating kilohertz QPO frequencies with Keplerian
frequencies at the disk-magnetosphere boundary. The mode frequencies to be
associated with upper or lower kilohertz QPOs are of the form
$\kappa (r) \pm m \Omega (r)$ where $m$ is an integer and the
epicyclic frequency $\kappa (r)$ is defined through $\kappa^2 \equiv
2\Omega ( 2\Omega + r d\Omega/dr )$. The rotation rate $\Omega(r)$
and the epicyclic frequency $\kappa (r)$ depend on the radial
position $r$ within the boundary layer \citep{EPA08}. The
frequencies and widths of the QPOs are likely to be affected by the
regions of the boundary layer that are dominant in the modulation of
the accretion flow to the neutron star. 
The observation of a correlation between the lower kilohertz QPO
frequency and the rate of change of this frequency with time
\citep{Sannaetal2012} provides direct evidence for the non-Keplerian
nature of the rotation rate $\Omega(r)$ in the boundary layer with
the existence of a maximum $\Omega(r)$. It is nice and rather rare
that such a simple analytical property of a model is verified by an
astrophysical observation.
\section{Non-Keplerian flow in the boundary layer is indicated by observations}
\citet{Sannaetal2012} have measured, for the first time, short time
scale time derivatives of the lower kilohertz QPO) frequency from the neutron-star
low-mass X-ray binary
4U 1636 - 53. They measure positive and negative time derivatives as
the QPO frequency wanders. The data exhibit a remarkable, simple correlation, that
the absolute value of the time derivative of
the lower kilohertz QPO frequency decreases as that QPO frequency increases:
\begin{equation}
\frac{d \lvert\dot{\nu}\rvert}{d\nu} < 0.
\end{equation}
Following \citet{Sannaetal2012}, I will employ frequencies $\nu$ in Hz.
If the observed kilohertz QPO frequencies are mode
frequencies $\nu(r)$ at some radial positions $r$ in the boundary layer,
applying the chain rule, and using $dr/dt = V_r$ , the radial velocity of
matter in the boundary layer, Eq. (1) leads to
\begin{eqnarray}
\frac{d \lvert\dot{\nu}\rvert}{d\nu} & = & \frac{d}{d\nu} \lvert{\frac{d\nu}{dr}}\rvert \lvert{V_r}\rvert  \nonumber \\
& = & \frac{dr}{d\nu} \frac{d}{dr} \lvert{\frac{d\nu}{dr}}\rvert \lvert{V_r}\rvert  \nonumber \\
& = & \lvert{V_r}\rvert \; \lvert{\frac{dr}{d\nu}}\rvert \; \frac{d^2\nu}{dr^2} < 0.
\end{eqnarray}
The radial velocity $V_r$ is assumed to be uniform throughout the boundary layer, 
and  constant on the short time scales of variation of the kilohertz QPO frequencies, so that additional terms in Eq. (2) representing radial and temporal variations of $V_r$ are neglected. In the \citet{ShakuraSunyaev73} model $V_r$ depends on the mass-inflow rate only to the power $1/3$; $V_r \propto \dot{M}_{inflow}^{1/3}$, as noted by \citet{Sannaetal2012}. Fluctuations in the mass-inflow rate $\dot{M}_{inflow}$ from the companion are likely to be smoothed by the viscous transport through the disk. 
Eq. (2) shows that the run of rotation rate $\nu (r) $ in the boundary layer has a
negative second derivative,
\begin{equation}
\frac{d^2\nu}{dr^2} < 0.
\end{equation}
This means that $\nu (r)$ has a continuous derivative going through
zero at a smooth maximum at some finite radius $r_0$ in the boundary
layer. All of the choices for $\nu_{QPO}$ as a mode frequency in the
boundary layer, like ( $1/(2\pi)$ times) $\kappa(r)$ , $\Omega(r)$,
$\kappa(r) \pm \Omega(r)$ indeed have such a maximum in the boundary
layer. In fact this is the {\em characteristic} property of boundary layer modes
as pointed out by \citet{AlparPsaltis08}. QPO models employing general relativistic frequencies \citep{StellaVietri98} and resonances (e.g. \citet{Kluzniak etal04}) also have frequency maxima at radii 
around $r_{ISCO}$, the innermost stable circular orbit. General relativistic effects will fold in and play a secondary role to viscous and hydrodynamic effects leading to the non-keplerian flow in the boundary layer.
If the lower kilohertz QPO frequency were a Kepler frequency,
or were related to the upper kilohertz QPO frequency with a linear
relation as discussed by \citet{Sannaetal2012} and the upper kilohertz QPO were
Keplerian, then we would have
\begin{equation}
\lvert\dot{\nu}\rvert \propto \lvert{\frac{d\nu_K}{dr}}\rvert V_r =  \frac{3}{2} \frac{\nu_K}{r} V_r ,
\end{equation}
which increases as the QPO frequency increases, contrary to the
correlation observed by \citet{Sannaetal2012}. Substituting measured values 
of $\lvert\dot{\nu}\rvert$ and assuming 
$\nu_{QPO} \cong \nu_K$ gives a very low value of the
radial flow rate $V_r$ in the boundary layer, requiring a very low
value of the Shakura-Sunyaev thin disk $\alpha$ parameter, $\alpha
\sim 10^{-6}$ \citep{Sannaetal2012}. Associating the kilohertz QPO frequencies 
with mode frequencies in the boundary layer solves both of these problems and
explains the results of \citet{Sannaetal2012} naturally. If the lower kHz QPO frequency is 
either $\kappa$ or $\kappa -\Omega$,
\begin{equation}
\lvert\dot{\nu}\rvert = \frac{1}{2\pi}[\frac{\partial\kappa}{\partial\Omega},
\; or \;(\frac{\partial\kappa}{\partial\Omega} - 1)]\lvert{\frac{\partial\Omega}{\partial r}}\rvert V_r .
\end{equation}
The frequency band is determined by the location near the maximum of
$\Omega$ and $\kappa$ of those parts of the boundary region that
predominate in the formation of the QPO signal. Here
$\lvert{\frac{\partial\Omega}{\partial r}}\rvert$ and therefore
$\lvert\dot{\nu}\rvert$ do decrease as the frequency increases
towards the maximum. As $\lvert{\frac{\partial\Omega}{\partial
r}}\rvert$ is small and
$[\frac{\partial\kappa}{\partial\Omega}, \; or
\;(\frac{\partial\kappa}{\partial\Omega} - 1)] $ are of O(1), $V_r$
and $\alpha$ are not unduly small.
\section{Discussion and Conclusions}
In models involving the disk boundary, the QPO signals in the accretion luminosity 
are due to variations in the flow 
through the boundary layer that may be intrinsic or excited through resonant interactions with the neutron star. These variations are wave-packets 
of the boundary layer normal modes visualized as ''blobs".  The frequencies and growth or decay time-scales of the boundary layer normal modes are calculated by modelling the local dynamics at each radial position $r$. The growth-and decay times  are likely to be much longer than the $\sim 100 s$ time scales of the observations analyzed by \citet{Sannaetal2012} if the sound speed $c_s < 0.5 \Omega r$ for $\alpha < 0.01$, or $c_s < 0.2 \Omega r$ if $\alpha = 0.1$ \citep{EPA08}.
Wave-packets that effect the accretion with 
discernibly high quality QPOs are likely to arise in the parts of the boundary layer where the frequency
spread $\delta \nu = \partial \nu / \partial r \delta r$ is relatively small, i.e. near the 
mode frequency maximum. The boundary layer is in interaction with the neutron star magnetosphere so that its excitations are imprinted on the accretion flow and show up as QPOs in the accretion luminosity. With this reasonable assumption, as the wave-packet moves through the boundary layer with speed $V_r$ the QPO frequency will have a derivative as employed in the derivation leading to Eq. (2). The QPO frequency derivative is positive or negative, with similar absolute values as observed by \citet{Sannaetal2012}, depending on the sign of $\partial \nu / \partial r$, positive when the wave-packet is closer to the star than the location of the maximum, and negative when the wave-packet is beyond that location. Fig.2 of \citet{Sannaetal2012} shows the frequency spread of the lower kilohertz QPO decreases to about 2.5 Hz at a QPO frequency of about 920 Hz. Fig.3 shows that the absolute value of the QPO frequency derivative decreases to about a fifth of its value at 650 Hz. The few frequency derivative measurements just above 920 Hz have larger error bars, and the frequency spread and its errors are also larger. These trends are suggestive that the maximum mode frequency associated with the lower kilohertz QPOs may be about 920 Hz. 
The presence of frequency maxima may hold clues to other QPO phenomenology. For example, a 3:2 ratio of frequencies corresponds to the $\kappa + \Omega$ and $\kappa$ modes near their maxima, where $\kappa = 2\Omega$ (see \citet{Erkut11} for an application to black hole QPO). Parallel QPO frequency-count rate tracks \citep{Mendez etal99} may correspond to runs of mode frequencies in the boundary layer at epochs of different $<\dot{M}_{inflow}>$ \citep{vdKlis01} with different inner disk radii and maximum mode frequencies.
The observation that the absolute value of the time derivative of
the lower kilohertz QPO frequency decreases as that QPO frequency
increases, provides direct evidence, within models that associate
the kilohertz QPOs with modes of the accretion disk boundary layer,
of the existence of a maximum in the non-Keplerian rotation rate
$\Omega(r)$ in the boundary layer. The boundary layer rotation curve
is expected to have a maximum as it evolves from the Keplerian
rotation curve in the disk to match the ("slow rotator") neutron
star's rotation rate $\Omega_{*}$ at the magnetospheric boundary.
The astrophysical observation of \citet{Sannaetal2012} verifies this
simple analytical property of disk boundary layer models.
\section*{Acknowledgments}
I thank Hakan Erkut and Dimitrios Psaltis for useful conversations, and the referee for helpful comments. This author is a member of the Science Academy, Bilim Akademisi, Turkey.





\begin{thebibliography}{}
\bibitem[Alpar \& Psaltis (2008)]{AlparPsaltis08} Alpar, M.A. \& Psaltis, D., 2008, MNRAS, 391, 1472
\bibitem[Alpar \& Shaham (1985)]{AlparShaham85} Alpar, M.A. \& Shaham, J. 1985, Nature, 316, 239
\bibitem[Alpar et al. (1992)]{Alparetal92} Alpar, M.A., Hasinger, G., Shaham, J. \& Yancopoulos, S., 1992, A\&A, 257, 627
\bibitem[Alpar \& Y{\i}lmaz (1997)]{AY97} Alpar M. A., Y{\i}lmaz A., 1997, New Astron., 2, 225
\bibitem[Campana (2000)]{Campana00} Campana, S., 2000, ApJ, 534, L79
\bibitem[Cui (2000)]{Cui00} Cui, W., 2000, ApJ, 534, L31
\bibitem[Erkut (2011)]{Erkut11} Erkut, M.H., 2011, ApJ, 743, 5
\bibitem[Erkut \& Alpar (2004)]{EA04} Erkut, M.H. \& Alpar, M.A. 2004, ApJ, 617, 461
\bibitem[Erkut, Psaltis \& Alpar (2008)]{EPA08} Erkut, M.H., Psaltis, D. \& Alpar, M.A. 2008, ApJ, 687,1220
\bibitem[Kato (2001)]{Kato01} Kato, S. 2001, PASJ, 53, 1
\bibitem[Kato (2009)]{Kato09} Kato, S. 2009, PASJ, 61, 1237
\bibitem[Kato (2012)]{Kato12} Kato, S. 2012, PASJ, 64, 129
\bibitem[Kluzniak et al. (2004)]{Kluzniak etal04} Kluzniak, W., Abramowicz, M.A. , Kato, S., Lee, W.H. \& 
Stergioulas, N. 2004, ApJ, 603, L89
\bibitem[Mendez et al. (1999)]{Mendez etal99} Mendez, M., van der Klis, M., Ford, E.C., Wijnands, R. \& van Paradijs, J. 1999, ApJ, 511, L49
\bibitem[Sanna et al. (2012)]{Sannaetal2012} Sanna, A., Mendez, M., Belloni, T. \& Altamirano, D. 2012, MNRAS, 424, 2936
\bibitem[Shakura \& Sunyaev (1973)]{ShakuraSunyaev73} Shakura, N.I. \& Sunyaev, R.A. 1973, A{\&}A, 24, 337 
\bibitem[Stella \& Vietri (1998)]{StellaVietri98} Stella, L. \& Vietri, M. 1998, ApJ, 492, L59 
\bibitem[Strohmayer et al. (1996)]{Strohmayeretal96} Strohmayer, T.E., Zhang, W., Swank, J.H., Smale, A., Titarchuk, L. \& Day, C. 1996, ApJ, 469, L9
\bibitem[van der Klis et al. (1985)]{vdKlisetal85} van der Klis, M., Jansen, F., van Paradijs, J., Lewin, W.~H.~G., van den Heuvel E.~P.~J., Tr\"{u}mper, J.~E., \& Sztajno, M.,  1985, Nature, 316, 225
\bibitem[van der Klis et al. (1996)]{vdKlisetal96} van der Klis, M., Swank, J.H., Zhang, W., Jahoda, K., Morgan, E.H., Lewin, W.H.G., Vaughan, B. \& van Paradijs, J. 1996, ApJ, 469, L1
\bibitem[van der Klis (2000)]{revvdKlis00} van der Klis, M. 2000, ARA{\&}A 38, 717
\bibitem[van der Klis (2001)]{vdKlis01} van der Klis, M. 2001, ApJ, 561, 943
\bibitem[van der Klis (2010)]{revvdKlis10} van der Klis, M. 2010, in ``Compact Stellar X-ray Sources", eds. W. Lewin  \& M. van der Klis, Cambridge University Press
\bibitem[Wagoner (1999)]{Wagoner99} Wagoner, R.~W., 1999, Phys. Rep., 311, 259
\end{thebibliography}



\bsp    
\label{lastpage}
\end{document}